\documentstyle[aps,epsfig,floats]{revtex}

\draft
\def\spose#1{\hbox to 0pt{#1\hss}}
\def\lta{\mathrel{\spose{\lower 3pt\hbox{$\mathchar"218$}}
     \raise 2.0pt\hbox{$\mathchar"13C$}}}
\def\gta{\mathrel{\spose{\lower 3pt\hbox{$\mathchar"218$}}
     \raise 2.0pt\hbox{$\mathchar"13E$}}}
\input epsf  
\newcommand{\lbfig}[1]{\refstepcounter{fig} \label{#1} }
\newcounter{fig}

\begin{document}

\twocolumn[\hsize\textwidth\columnwidth\hsize\csname
@twocolumnfalse\endcsname \title{Cosmological Consequences of
Slow-Moving Bubbles in First-Order Phase Transitions}

\author{Anne-Christine Davis\footnotemark \ and Matthew Lilley\footnotemark}

\address{Department of Applied Mathematics and Theoretical Physics,
  University of Cambridge,\\ Cambridge CB3~9EW, United Kingdom }

\date{\today}
\maketitle

\pagenumbering{arabic}

\begin{abstract}In cosmological first-order phase transitions, the
  progress of true-vacuum bubbles is expected to be significantly
  retarded by the interaction between the bubble wall and the hot
  plasma.  We examine the evolution and collision of slow-moving
  true-vacuum bubbles.  Our lattice simulations indicate that phase
  oscillations, predicted and observed in systems with a local
  symmetry and with a global symmetry where the bubbles move at speeds
  less than the speed of light, do not occur inside collisions of
  slow-moving local-symmetry bubbles.  We observe almost instantaneous
  phase equilibration which would lead to a decrease in the expected
  initial defect density, or possibly prevent defects from forming at
  all. We illustrate our findings with an example of defect formation
  suppressed in slow-moving bubbles.  Slow-moving bubble walls also
  prevent the formation of `extra defects', and in the presence of
  plasma conductivity may lead to an increase in the magnitude of any
  primordial magnetic field formed.
\end{abstract}
]

\footnotetext{~\vspace{-.8cm}~} 
\footnotetext{\footnotemark[2] Electronic address: {\tt
    A.C.Davis@damtp.cam.ac.uk}}
\footnotetext{\footnotemark[1]
  Electronic address: {\tt M.J.Lilley@damtp.cam.ac.uk}}

\renewcommand{\thefootnote}{\arabic{footnote}}


\section{Introduction}

According to standard cosmology, the early universe is expected to
have undergone a series of symmetry-breaking phase transitions as it
expanded and cooled, at which topological defects may have formed
\cite{shellard_vilenkin}.  Phase transitions are labelled first- or
second-order, according to whether the position of the vacuum state in
field space changes discontinuously or continuously as the critical
temperature is crossed.  A first-order phase transition proceeds by
bubble nucleation and expansion. When at least $(4-n)$ of these
bubbles collide (for $n=0, 1$ or $2$), an $n$-dimensional topological
defect may form in the region between them.

In recent years there has been considerable interest in the formation
of defects in first-order phase transitions, in particular the
validity of the so-called geodesic rule.  The geodesic rule, first
stated by Kibble \cite{kibble}, predicts that after a two-bubble
collision the phase of the scalar field interpolates continuously
between the values in each bubble, along the shortest path in field
space.  Early analysis \cite{HDB} confirmed the geodesic rule
for defect formation in both global and local theories, albeit using a
planar approximation and neglecting the effect of the surrounding
plasma.  In later work, the finite conductivity of the plasma was
considered \cite{kibble_vilenkin} for local theories, as was the
effect of slow-moving (i.e.\ speeds less than the speed of light)
bubble walls in theories with a global symmetry \cite{ferrera_melfo}.
These analyses confirm defect formation in first-order phase
transitions, but make conflicting claims about the number of defects
actually formed.  In this paper we investigate this issue.  Unlike
previous work, we include the effect of slow-moving bubble walls in
both the global {\em and} local cases. We use our results to make
qualitative comparisons between defect densities formed in global and
local theories, and by slow-moving and fast-moving bubble walls.  As
well as the consequences for defect formation, we also consider the
implications of slow-moving walls for the formation of primordial
magnetic fields at a first-order phase transition.
  
We take as our model the simplest spontaneously-broken gauge symmetry:
the Abelian Higgs model, which has a local $U(1)$ symmetry, with
Lagrangian
\begin{equation} \label{ssbscalar}
{\mathcal L} =  {( {D}_{\mu} \Phi )}^{\dagger}{( {D}_{\mu} \Phi )} - {1 \over
  4}F_{\mu \nu}F^{\mu \nu}- V(\Phi^{\dagger}\Phi),
\end{equation}
where ${D}_{\mu} \Phi = { \partial }_{\mu}\Phi - ie{A}_{\mu}\Phi$ and
$F_{\mu \nu} = { \partial }_{\mu}{A}_{\nu} - { \partial
  }_{\nu}{A}_{\mu}$.  The detailed form of the effective potential
$V(|\Phi|)$ will depend upon the particular particle-physics model
being considered, but in order to be able to study the generic
features of a first-order phase transition we shall take $V$,
following Ferrera and Melfo \cite{ferrera_melfo}, to be
\begin{equation} \label{potentialeqn}
V(\Phi) = \lambda \Bigl[{\frac{|\Phi |^{2}}{2}}(|\Phi | -
  \eta)^{2} - \frac{\varepsilon}{3}{\eta}|\Phi |^{3}\Bigr],
\end{equation}
where in a realistic model, $\varepsilon = \varepsilon (T) \propto
(T_c - T)$.  $V$ has a local minimum false-vacuum state at $\Phi = 0$
which is invariant under the $U(1)$ symmetry, and global minima
true-vacuum states on the circle $|\Phi |= \rho_{tv} \left(\eta /
  4\right)(3 + \varepsilon + \sqrt{1 + 6\varepsilon + \varepsilon^2})$
which possess no symmetry.  The dimensionless parameter $\varepsilon$
is responsible for lifting the degeneracy between the two sets of
minima -- the greater $\varepsilon$, the greater the potential
difference between the false- and true-vacuum states, and hence the
faster the bubbles will accelerate.  By making the field and
coordinate transformations
\begin{eqnarray} \label{transformations}
\Phi \longrightarrow \phi &=& \eta  \Phi \\
{\bf x} \longrightarrow {\bf x'} &=& \frac{{\bf x}}{\sqrt{\lambda}\eta} \\
t \longrightarrow t' &=& \frac{t}{\sqrt{\lambda}\eta}
\end{eqnarray}
it is possible to set $\lambda$ and $\eta$ to unity, so that the
potential is parametrized only by $\varepsilon$, and hereafter we
shall use these transformed variables.

The bubble nucleation rate per unit time per unit volume is given by
the `bounce' solution of the Euclidean field theory \cite{coleman}.
Ignoring quantum fluctuations, the phase $\theta$ is constant within
each bubble, and uncorrelated between spatially-separated bubbles.
Any non-zero gauge fields in the nucleation configuration will make a
contribution to the action and hence the nucleation of bubbles with
non-zero gauge fields is exponentially suppressed.  When three or more
bubbles collide, a phase-winding of $2{\pi}n$ can occur around a
point, which by continuity must then be at $\Phi = 0$.  In three
spatial dimensions, this topologically-stable region of high-energy
false vacuum is string-like -- a cosmic string.  

The formation and evolution of cosmic strings have been studied in
great detail.  Cosmic strings have been evoked as, amongst other
things, possible seeds for cosmic structure formation, sources of
cosmic rays, gravitational radiation and baryogenesis (see, e.g.
\cite{shellard_vilenkin}).  In order to be able to assess the
significance of cosmic strings in the evolution of the early universe,
it is important to be able to estimate the initial defect density
accurately.  This depends on how the phases between two or more
bubbles interpolate after collision.  In particular, although strings
are in general formed when three or more bubbles collide, a
simultaneous three-bubble collision is unlikely -- one would expect in
general two-bubble collisions, with a third, or fourth bubble
colliding some finite time later.  If the phase inside a two-bubble
collision is able to equilibrate quickly, and before a third bubble
arrives, there may be a strong suppression of the initial string
density.  The effect of phase equilibration on the initial defect
density was first investigated by Melfo and Perivolaropoulos
\cite{MP}.  They found a decrease of less than $10\%$, in models which
possess a global symmetry and with bubbles moving at the speed of
light.

The above description of defect formation, however, ignores any effect
that the hot-plasma background may have on the evolution of the Higgs
field, which may be significant in the early universe.  Real-time
simulations \cite{laine} and analytic calculations \cite{bubble} for
the (Standard Model) electroweak phase transition predicted that the
bubble wall would reach a terminal velocity $v_{ter} \sim 0.1c$.  The
reason for this is simple: outside the bubble, where the ($SU(2)\times
U(1)$) symmetry remains unbroken, all fields coupled to the Higgs are
massless, acquiring their mass from the vacuum expectation value of
the Higgs in the spontaneously-broken symmetry phase inside the
bubble.  Particles outside the bubble without enough energy to become
massive inside bounce off of the bubble wall, retarding its progress
through the plasma.  The faster the bubble is moving, the greater the
momentum transfer in each collision, and hence the stronger the
retarding force.  Thus a force proportional to the bubble-wall
velocity appears in the effective equations of motion.

Ferrera and Melfo \cite{ferrera_melfo} studied bubble collisions in such
an environment, for theories which possessed a global symmetry, and
found that decaying phase oscillations occur inside a two-bubble
collision, leading to a suppression of the defect formation rate
\cite{ferrera98}.  Kibble and Vilenkin \cite{kibble_vilenkin} studied
phase dynamics in collisions of undamped bubbles in models with a
local symmetry, and found, analytically, a different kind of decaying
phase oscillation.  When the finite conductivity of the plasma was
included, these oscillations were found not to occur.  However, Kibble
and Vilenkin did not consider the behaviour of the phase after
collisions of bubbles moving at speeds slower than the speed of light.
Moreover, because of the symmetry assumptions made in their
calculations, their results cannot be simply extrapolated to the
slower-moving case.

The behaviour of the phase inside bubble collisions in local theories
where the bubbles move at the speed of light, and in global theories
with slow-moving bubbles has been considered.  However, the most
realistic scenario cosmologically -- a gauge-theory phase transition
where the bubbles are slowed significantly by the plasma (as might be
expected at the electroweak- or GUT-scales) -- has not been studied.
This paper presents the results of our investigations into what
happens in theories with a local symmetry, where the bubbles are
moving at terminal velocities less than the speed of light.  Our
$3+1$-dimensional simulations indicate that, for slow-moving bubbles,
phase oscillations of either of the types described in
\cite{ferrera_melfo} or \cite{kibble_vilenkin} do not occur, before the
effect of the plasma conductivity is even considered.  We therefore
expect that (a) {\em fewer} defects would form in a phase transition where
the `Higgs' field is coupled to a gauge field than in a
global-symmetry phase transition, and (b) in local
theories, {\em fewer} defects would form in slow-moving bubbles than
fast-moving ones.

We should note in passing that we have ignored the effect of the
expansion of the universe in our work.  This is a good approximation
for phase transitions which take place at late times, like the
electroweak phase transition.  At phase transitions which occur
earlier, however, the Hubble expansion may have a significant effect
on bubble and phase dynamics.  This topic deserves consideration on its own,
and work is currently in progress \cite{lilley}.

In the following section, we describe the effects of a slow-moving
bubble wall on phase dynamics inside bubbles collisions in theories
which possess a global symmetry.  In section III, we discuss phase
equilibration in theories with a local symmetry and present our new
results in the case of slowly-moving local-symmetry bubbles.  Our
conclusions are supported by examples of defect formation suppressed
in slow-moving bubbles.  We show that the `extra defects' found in
\cite{ed_paul} do not occur in heavily-damped environments.  In
section IV we discuss the formation of a primordial magnetic field.
We show that the presence of the plasma conductivity results in a
larger magnetic field for fast-moving bubbles. For slow-moving
bubbles, the plasma conductivity stops the field dispersing. A larger
magnetic field could also result in this case.  A discussion of our
results and conclusion are presented in section V.

\section{Global Symmetry}

If the gauge coupling $e$ is set to zero, we have a theory with a
global $U(1)$ symmetry
\begin{equation}\label{lagrangian}
{\mathcal L} = {( {\partial}_{\mu} \Phi )}^{\dagger}{(
  {\partial}_{\mu} \Phi)} - V(\Phi^{\dagger}\Phi).
\end{equation}
By writing $\Phi = \rho e^{i\theta}$, the equations of motion for
the modulus $\rho$ and phase $\theta$ of the Higgs field are
\begin{eqnarray} \label{rho_theta}
\ddot{\rho} - {\rho}'' - ({\partial}_{\mu}\theta)^{2}\rho & =
 &-\frac{\partial V}{\partial \rho}\\
\label{rho_theta2}
 \partial^{\mu}\bigl[\rho^{2}\partial_{\mu}\theta\bigr] &=& 0.
\end{eqnarray}
If the potential difference between the true- and false-vacuum states
is much smaller than the height of the barrier separating them, the
field equations may be solved using the `thin-wall' approximation
\cite{coleman}, by setting $\varepsilon=0$.  For our potential
(\ref{potentialeqn}), this yields
\begin{equation} \label{initial}
|\Phi| = \frac{\eta}{2}\left[1 +
  \tanh\left(\frac{\sqrt{\lambda}\eta}{2}\left(s -
  R_{0}\right)\right)\right],
\end{equation}
where $s^2 = {\bf x}^2 - t^2$ and $R_{0}$ is the bubble radius on
nucleation.

Note that $\theta = {\rm constant}$ trivially satisfies the phase
equation (\ref{rho_theta2}), and so if the phase is initially constant
within each bubble, as we shall assume, there are no phase dynamics
until the bubbles collide.

As described in the introduction however, we would like to investigate
the behaviour of the phase in collisions of slow-moving bubbles.  For
a given theory, by considering the Boltzmann equations for scattering
off of the Higgs field, it is possible to calculate the terminal
velocity of the bubble wall \cite{turok}.  Since we are not concerned
here with the parameters of a specific particle-physics model, we
choose instead to use a single damping parameter $\Gamma$ to model the
interaction of the Higgs with the plasma.  In the introduction we
claimed that the plasma would introduce a term proportional to the
bubble-wall velocity into the equations of motion.  Since the phase
$\theta$ of the Higgs field is not affected by the effects described,
we assume that the plasma couples only to the modulus $\rho$.  We then
have effective equations of motion
\begin{eqnarray}
\label{rho_damp} \ddot{\rho} - {\rho}''  + \Gamma \dot{\rho} -
({\partial}_{\mu}\theta)^{2}\rho & = & -\frac{\partial V}{\partial
 \rho}\\
 \label{theta_damp}
 \partial^{\mu}\bigl[\rho^{2}\partial_{\mu}\theta\bigr] &=& 0.
\end{eqnarray}
A damping term of this form has been used by several authors
\cite{ferrera_melfo}, \cite{heckler}, \cite{turok}, and has also been
derived from the stress-energy of the Higgs, assuming a coupling to
the plasma \cite{ignatius}.  Heckler \cite{heckler} estimates $\Gamma
\sim g_{W}^2 T_c$ for the electroweak phase transition, by comparing
the energy generated by the frictional damping with the pressure on
the wall due to the damping.

The effect of this damping term is that instead of accelerating up to
the speed of light, the bubble walls reach a terminal velocity
$v_{ter} < c$.  By making the {\em ansatz} $\rho =
\rho\left[x-x_0(t)\right]$, the terminal velocity can be calculated
\cite{ferrera_melfo} by integrating the equation of motion for $\rho$
\begin{equation}\label{vter}
v_{ter} = \frac{\Delta V}{\Gamma \int{\rho'^2dx}},
\end{equation}
where $\Delta V$ is the difference in potential energy between the
true- and false-vacuum states. Assuming that the wall has a
Lorentz-contracted, moving profile of the form
(\ref{initial}) \cite{ferrera_melfo}
\begin{equation}\label{damped_profile}
\rho = \frac{\rho_{tv}}{2}\left[1 +
 \tanh\left(\frac{\sqrt{\lambda}\rho_{tv} \gamma}{2}\left(r - v_{ter}t
 - R_0 \right)\right)\right],
\end{equation}
the integral in the denominator of (\ref{vter}) can be evaluated.
Expanding $\gamma = (1 - v_{ter}^2)^{-1/2}$, we obtain
\begin{equation}\label{vter_full}
v_{ter} = \frac{A}{\sqrt{A^2 + \Gamma^2}},
\end{equation}
where $A = {6 \, \Delta V / \sqrt{\lambda} \rho_{tv}^3}$.  We have
simulated the evolution of bubbles in such a dissipative environment
in $1+1$-dimensions.  Taking a static profile of the form
(\ref{initial}) as the initial conditions, the terminal velocity of
the bubble was calculated for a range of values of friction parameter
$\Gamma$.  The accuracy of formula (\ref{vter_full}), compared with
terminal velocities calculated directly from simulations can be seen
in Figure \ref{velocity}.  This is a very useful result, as from it we
can dial the input value of $\Gamma$ to produce the value of $v_{ter}$
corresponding to the particular particle-physics model we are
interested in.  Heckler \cite{heckler}, and Ferrera and Melfo
\cite{ferrera_melfo} obtained a result like (\ref{vter}), and Haas
\cite{haas} found the best-fit equation $v_{ter} = A +
\left(1-A\right)/\left(1 + B \Gamma^{1.62}\right)$ from
Langevin-equation simulations.  However, equation (\ref{vter_full}),
we believe, holds for all of the cases above, provided that
$\varepsilon$ is small enough for the `thin wall' approximation to
hold, and is more useful when performing simulations.  For example, it
could be applied to the electroweak phase transition in the
supersymmetric case, if the terminal velocity of the bubble wall were
calculated.

\begin{figure}[htb]
\begin{center}
\epsfig{figure=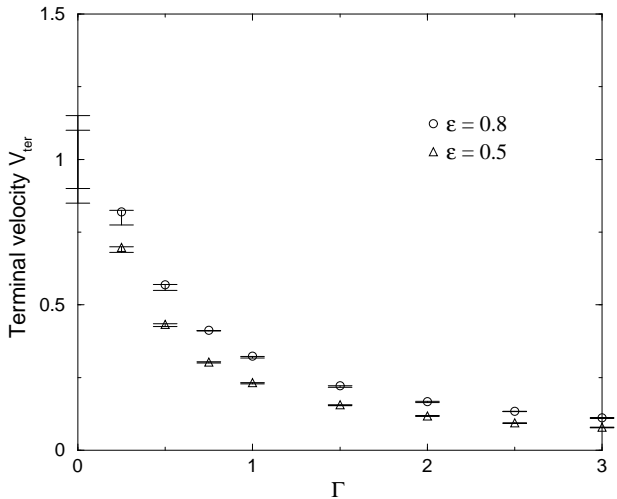, width=8cm}
\lbfig{velocity} 
\caption{Terminal velocity $v_{ter}$ of bubble walls vs. damping
  coefficient $\Gamma$ for $\varepsilon=0.8$ and $\varepsilon=0.5$.
  The error bars represent velocities calculated from simulations, and
  the circles and triangles the values obtained from
  (\ref{vter_full}).}
\end{center}

\end{figure}

Ferrera and Melfo \cite{ferrera_melfo} described how, in the context
of a theory with a global symmetry, slow-moving bubble walls lead to
phase oscillations.  When two bubbles collide, the walls merge.
Across the plane (in 3 spatial dimensions) of intersection, there
exists a phase gradient to drive equation (\ref{rho_theta2}), and a
phase wave propagates into each bubble from the centre -- see Figures
\ref{bounce} (b) and \ref{merge} (b).  As the Goldstone boson is
massless, and undamped, this wave travels at the speed of light.  If
the phase difference between the bubbles is $\Delta\theta$, the phase
wave will carry a phase difference $+\Delta\theta/2$ into one of the
bubbles, and $-\Delta\theta/2$ into the other, equilibrating the
phase.  If the bubble wall is moving at a terminal velocity $v_{ter} <
c$, the wave will catch up with the bubble wall, and rebound -- the
returning wave will now `flip' the original phase profile.  Thus phase
oscillations occur inside the merged bubbles.  Given three or more
spatially-separated bubbles whose distribution of phases one would
expect to generate a vortex on collision, a vortex, an anti-vortex, or
none at all may form, depending on the profile of the phase inside the
two bubbles at the moment of collision of the third -- an example of
how phase dynamics can affect the defect-formation process.  The
oscillations are damped, because the bubble walls continue to expand,
increasing the volume over which the finite-energy wave must sweep,
thus diluting the phase difference carried by the wave. Thus the
converse of the above statement is not true -- an initial distribution
of phases which one would not expect to form a defect, will {\em not}
produce one as a result of phase oscillations.  Statistical
simulations in two dimensions \cite{ferrera98} have shown that this
leads to a suppression in the defect-formation rate -- the slower the
bubble walls, the fewer defects are formed per nucleated bubble.

\begin{figure*}[ht]
\begin{center}
  \epsfig{figure=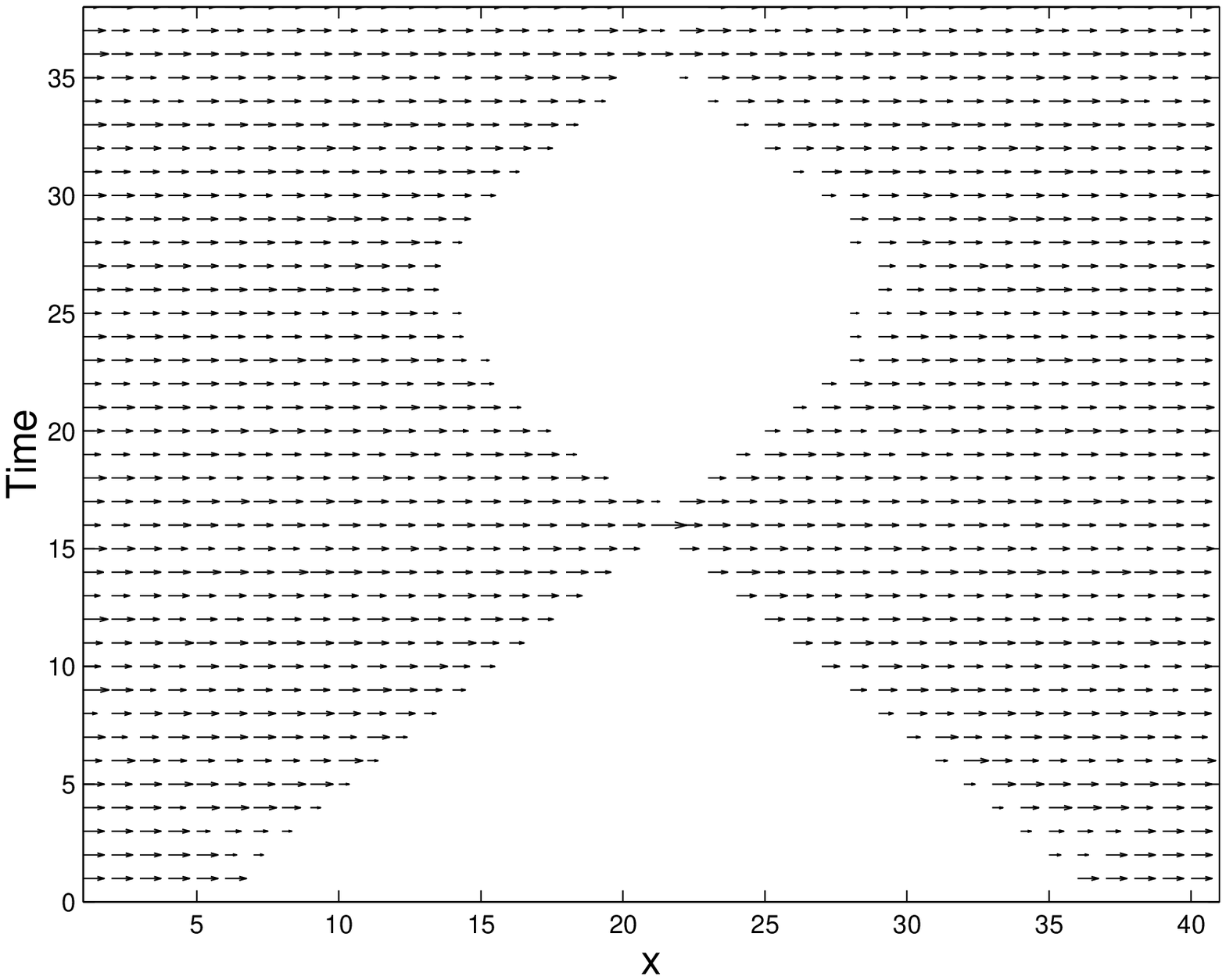,
  width=7.8cm}\hspace*{5mm}\epsfig{figure=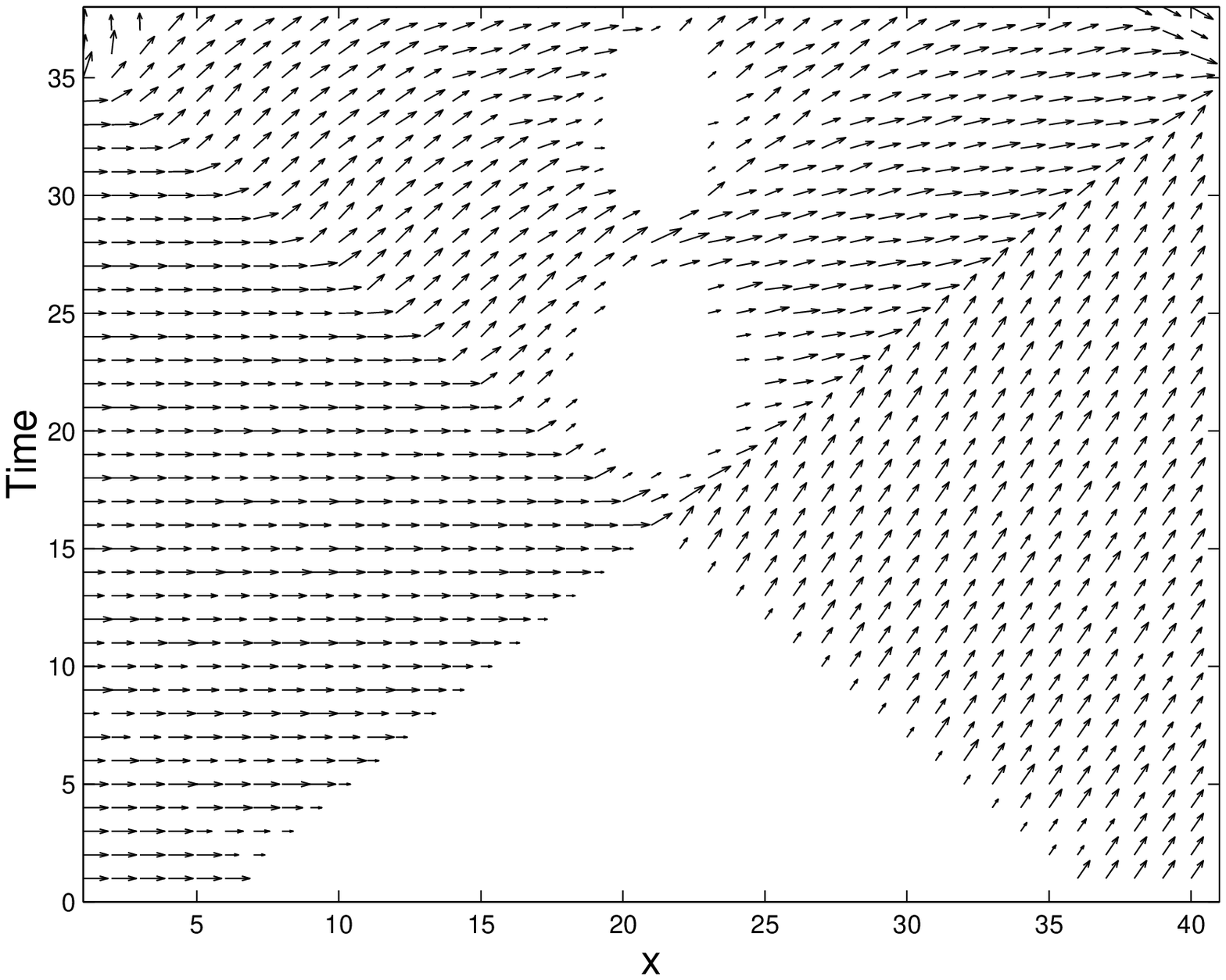, width=7.8cm}
  \lbfig{bounce}
\end{center}
\vspace*{-6.4cm}\noindent
\hspace*{0.6cm}{\large\bf a)} \hspace*{7.8cm}{\large\bf b)}
\vspace*{5.4cm}
\begin{center}
\caption{Undamped bubbles passing through one another: a) with no
phase difference, and b) with phase difference $\Delta \theta = \pi /
3$.  In both cases $\varepsilon = 0.1$. The direction of the arrows
gives the phase $\theta$ and their length is proportional to the
modulus $\rho$.  Note in case b) that a phase wave propagates at the
speed of light from the point of collision into the bubble interiors.
In this case, the symmetry-restored region is smaller than in a),
because the phase wave is able to carry away energy from the collision
point, which is not possible in case a), where there is no phase
difference between the two bubbles.}
\end{center}
\end{figure*}
\begin{figure*}[ht]
\begin{center}
  \epsfig{figure=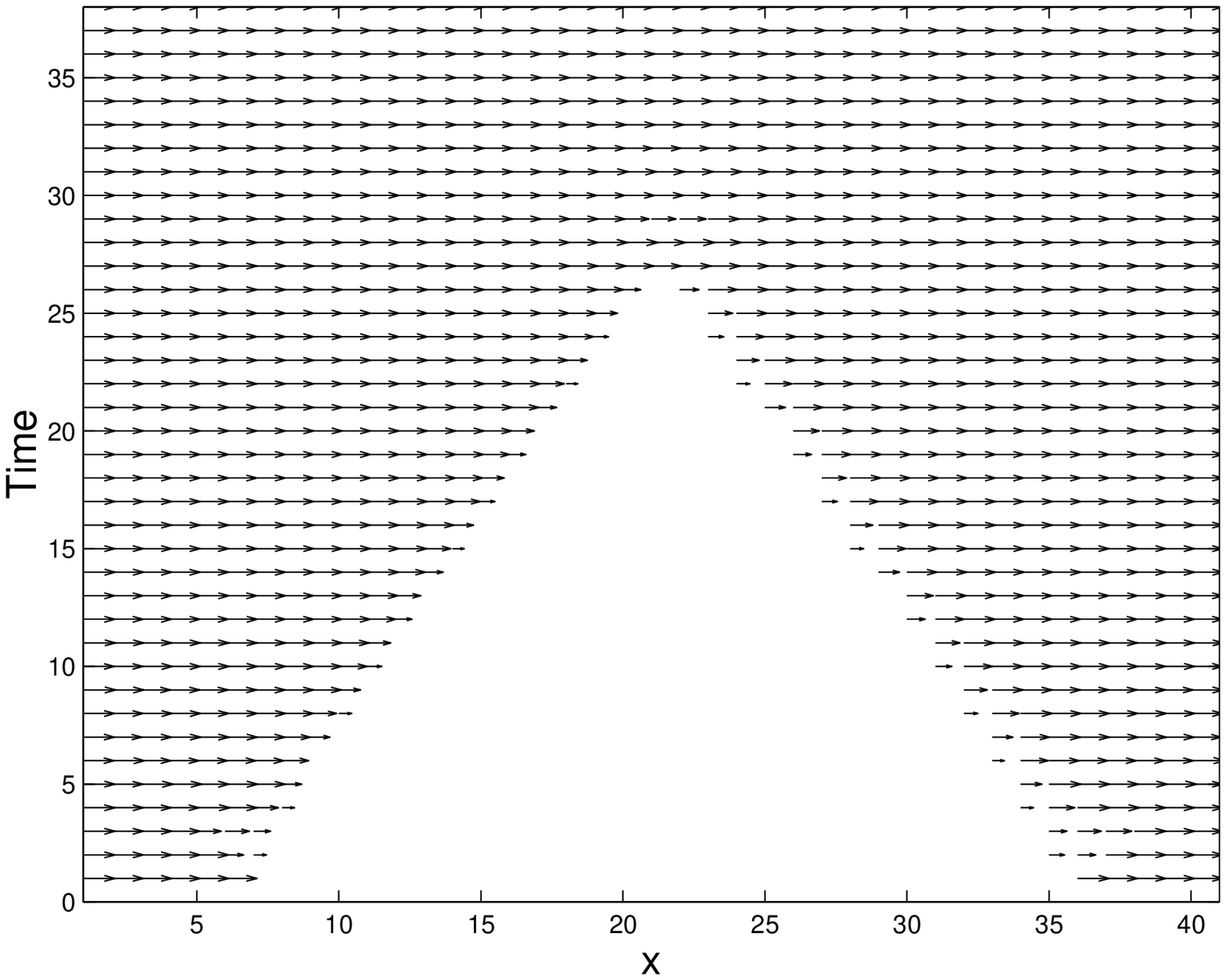,
  width=7.8cm}\hspace*{5mm}\epsfig{figure=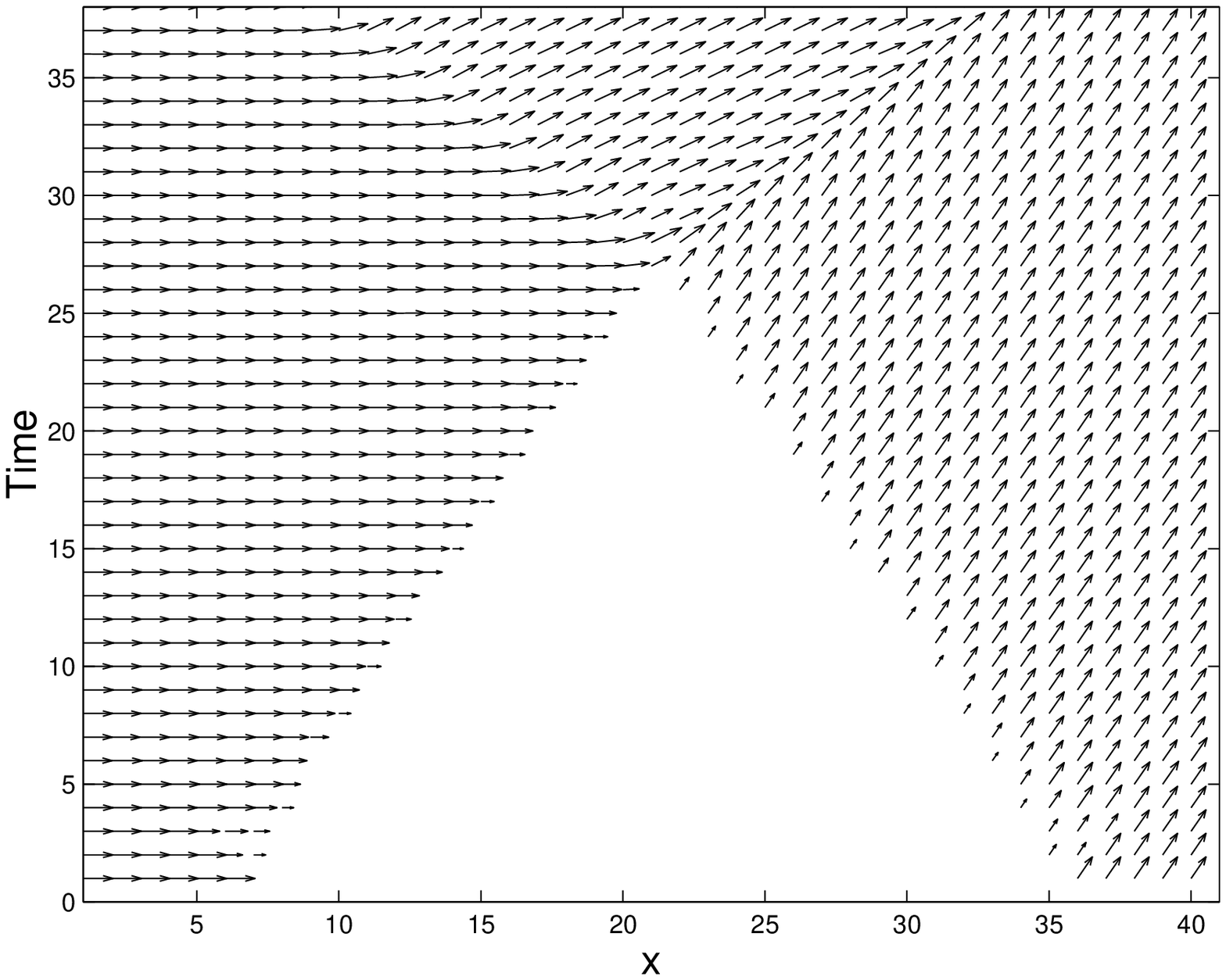, width=7.8cm}
  \lbfig{merge}
\end{center}
\vspace*{-6.4cm}\noindent
\hspace*{0.6cm}{\large\bf a)} \hspace*{7.8cm}{\large\bf b)}
\vspace*{5.4cm}
\begin{center}
\caption{Damped bubbles bubbles merging: a) with no phase difference,
and b) with phase difference $\Delta \theta = \pi / 3$.  In both cases
$\varepsilon = 0.1$.  Note in case b) the phase wave propagating
inside the bubbles at the speed of light after collision, faster than
the bubble walls.  This phase wave will catch up with the outer wall
of the bubble (not shown here), and rebound, producing phase
oscillations.}
\end{center}
\end{figure*}

\section{Local Symmetry}

\subsection{Phase dynamics inside two-bubble collisions}

Including the gauge fields in our model, the field equations become
\begin{eqnarray}
\label{local_rho} \ddot{\rho} - {\rho}'' - ({\partial}_{\mu}\theta -
  e{A}_{\mu})^{2}\rho & = &-\frac{\partial V}{\partial \rho}\\
\label{local_theta} \partial^{\mu}\bigl[\rho^{2}(\partial_{\mu}\theta
- eA_{\mu})\bigr] &=& 0\\ 
\label{local_a}\ddot{A_\nu} - {A_\nu}'' -
{\partial}_{\nu}\left(\partial
 \cdot A \right) & = & -2e \rho^2 {\partial}_{\nu}\theta.
\end{eqnarray}

Since we now have a local $U(1)$ symmetry, the phase $\theta$ can be
arbitrarily re-defined at any point in time by a gauge transformation,
and so we need a gauge-invariant notion of phase.  We define,
following Kibble and Vilenkin \cite{kibble_vilenkin}, the
gauge-invariant phase difference between two points $A$ and $B$
\begin{equation} \label{gipd}
\Delta \theta = \int_{A}^{B} dx^{i}\left(\partial_{i} - i e A_{i}\right),
\end{equation}
where $i=1,2,3$ and the integral is taken, for simplicity, along the
straight line joining $A$ and $B$.

For bubbles which move at approximately the speed of light, it is
possible to greatly simplify the field equations.  If we consider a
two-bubble collision, in a frame where the bubbles are nucleated
simultaneously, by assuming that the bubbles instantly propagate at
the speed of light, it is possible to impose $SO(1,2)$ Lorentz
symmetry on the field equations.  Thus the fields are functions of $z$
and $\tau^2 = t^2 - x^2 - y^2$ only.  With this assumption, and a
step-function {\em ansatz} for the phase $\theta$ at the time of
collision, Kibble and Vilenkin \cite{kibble_vilenkin} solved the field
equations for $\Delta \theta$
\begin{equation} \label{kvgipd}
\Delta \theta = \frac{2 R}{t} \theta_{0} \left( \cos e\eta\left(t - R\right)
    + \frac{1}{e \eta R}\sin e \eta\left(t - R\right)\right),
\end{equation}
where $2\theta_0$ is the initial phase difference between the
spatially-separated bubbles and $R$ is their radius on collision at
$t=0$.

Equation (\ref{kvgipd}) describes decaying phase oscillations, the
time scale of equilibration determined by the initial phase difference
and radius of the colliding bubbles, the frequency of oscillation by
the gauge-boson mass. These oscillations, and the accuracy of this
formula for small initial phase differences, have recently been
confirmed in simulations \cite{CST}.

However, the assumption that the bubbles move at, or close to the
speed of light, does not appear to be realistic \cite{bubble}.  In
this case, the symmetry assumptions made in \cite{kibble_vilenkin} are
no longer valid.  Moreover, it is not possible to replace the
coordinate $\tau^2 = t^2 - x^2 - y^2$ by the obvious choice $\tau'^2 =
\left(v_{ter}t\right)^2 - x^2 - y^2$, since only the the modulus of
the Higgs field, the bubble wall, is constrained in this way -- the
phase and gauge fields are still free to propagate causally.

In order to investigate whether phase oscillations -- which occur in
the global theory with slow-moving bubbles, and in the local theory
with fast-moving bubbles -- still occur in the local theory when the
bubbles expand slower than the speed of light, we include the
dissipation term $\Gamma \dot{\rho}$ into the equation for the modulus
of the Higgs field $\rho$, without coupling it to the phase or the
gauge fields.  This is motivated in the same manner as described in
the global case.  A term proportional to $\dot{\rho}$ is, of course,
$U(1)$ gauge-invariant.

Since there is no longer an obvious simplification of the equations of
motion which might lead to an analytic solution, we turn to computer
simulations. The equations of motion were discretized in the
gauge-invariant way described in \cite{mmr}, choosing the temporal
gauge $A_0 \equiv 0$ in order to make the time evolution trivial.  We
used a lattice of size $200^3$ and a lattice spacing $a=0.5$ -- tests
were performed on lattices with spacing down to $a = 0.1$ giving no
qualitatively-different results.  The time evolution was performed
using a fourth-order Runge-Kutta algorithm.  We took as initial
conditions a static profile of the form (\ref{initial}) for $\rho =
|\Phi|$, for two bubbles of radius $R=5$, with phases $\theta = 0$ and
$\theta = 2\pi / 3$, centred at $(\pm 8,0,0)$.  We choose to ignore
any primordial magnetic field and, since the nucleation process is not
expected to generate non-zero gauge fields (see Introduction), set all
the gauge fields to zero initially.

The results of the simulations are displayed in Figures \ref{gamma},
\ref{sigma} and \ref{vortex}.  For the sake of clarity and to aid
comparison between the different cases, we have chosen to present our
results in terms of the evolution with time of the gauge-invariant
phase difference $\Delta \theta$.  We evaluated $\Delta \theta$
between the centres of the two bubbles, though the qualitative
behaviour was found not to change when it was calculated between
different points.

Figure \ref{gamma} (a) shows the behaviour of the gauge-invariant
phase difference for bubbles moving at the speed of light -- the
decaying oscillations calculated by Kibble and Vilenkin in the local
case.  In the global case, $e=0$, we find that the phase {\em does}
equilibrate, but on a much longer time-scale.  Thus we would expect
that for fast-moving bubbles, fewer defects are formed in local
theories than global ones, since in order to form a defect a phase
difference inside the two merged bubbles must be present when a third
bubble collides.

In Figure \ref{gamma} (b) we plot $\Delta \theta$ for slower-moving
bubbles.  For $e=0$, we confirm in $3+1$-dimensions the decaying phase
oscillations described by Ferrera and Melfo \cite{ferrera_melfo} and
observed by them in $2+1$-dimensions.  These oscillations are killed
by adding in gauge fields -- for a fixed bubble-wall velocity, the
stronger the gauge coupling, the less time the gauge-invariant phase
difference is non-zero, and hence the less likely a third collision
will occur in time for a defect to form.  Thus we would expect a lower
defect-formation rate in local theories with slower-moving bubble
walls.
\begin{figure*}[ht]
\begin{center}
  \epsfig{figure=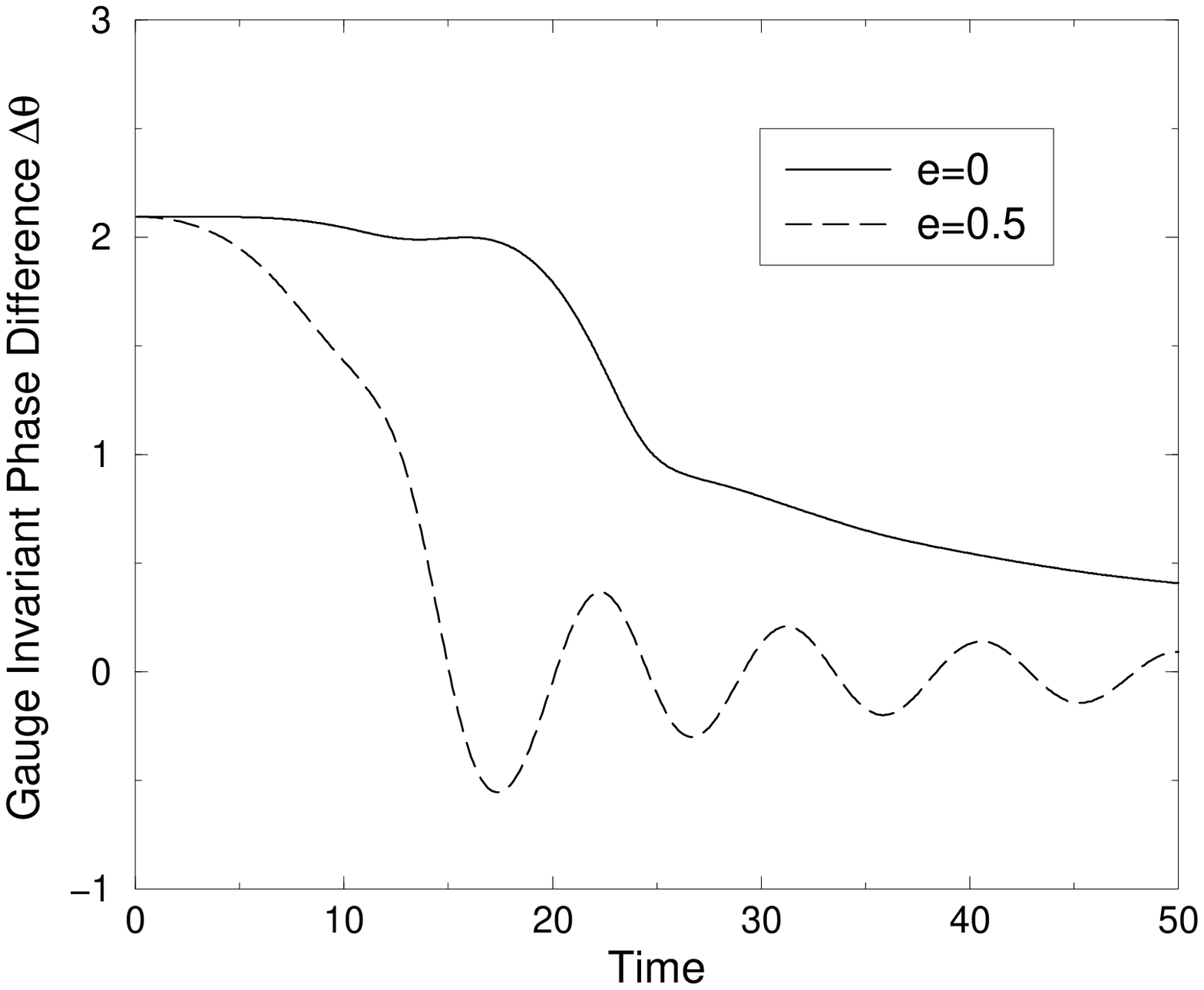,
  width=7.8cm}\hspace*{5mm}\epsfig{figure=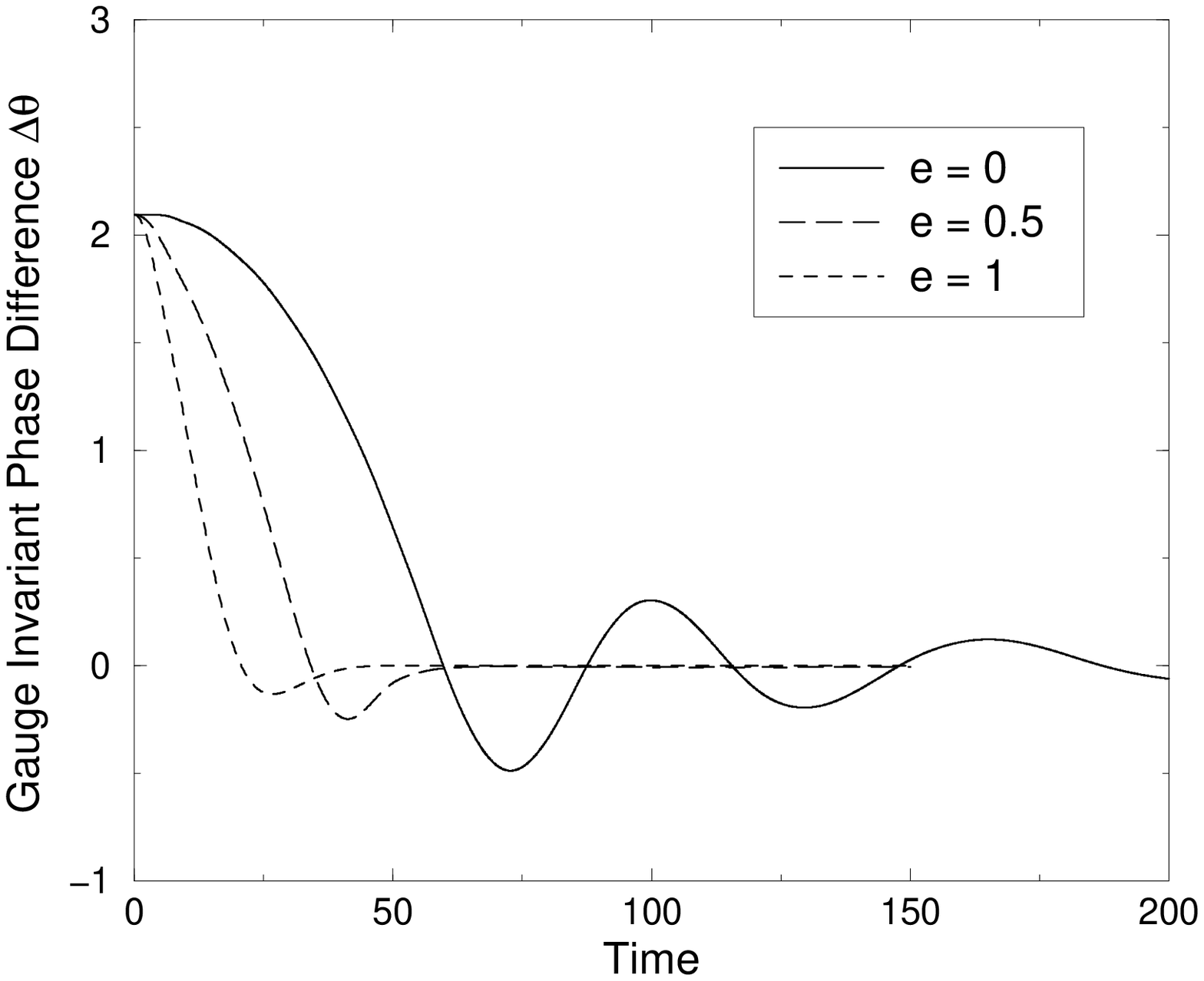, width=7.8cm}
  \lbfig{gamma}
\end{center}
\vspace*{-6.4cm}\noindent
\hspace*{1.9cm}{\large\bf a)} \hspace*{7.8cm}{\large\bf b)}
\vspace*{5.4cm}
\begin{center}
\caption{Gauge-invariant phase difference between a) two bubbles moving
  at the speed of light , $\Gamma = 0$, and b) two slow-moving
  bubbles, $\Gamma = 2$.}
\end{center}
\end{figure*}
\begin{figure*}[ht]
\begin{center}
  \epsfig{figure=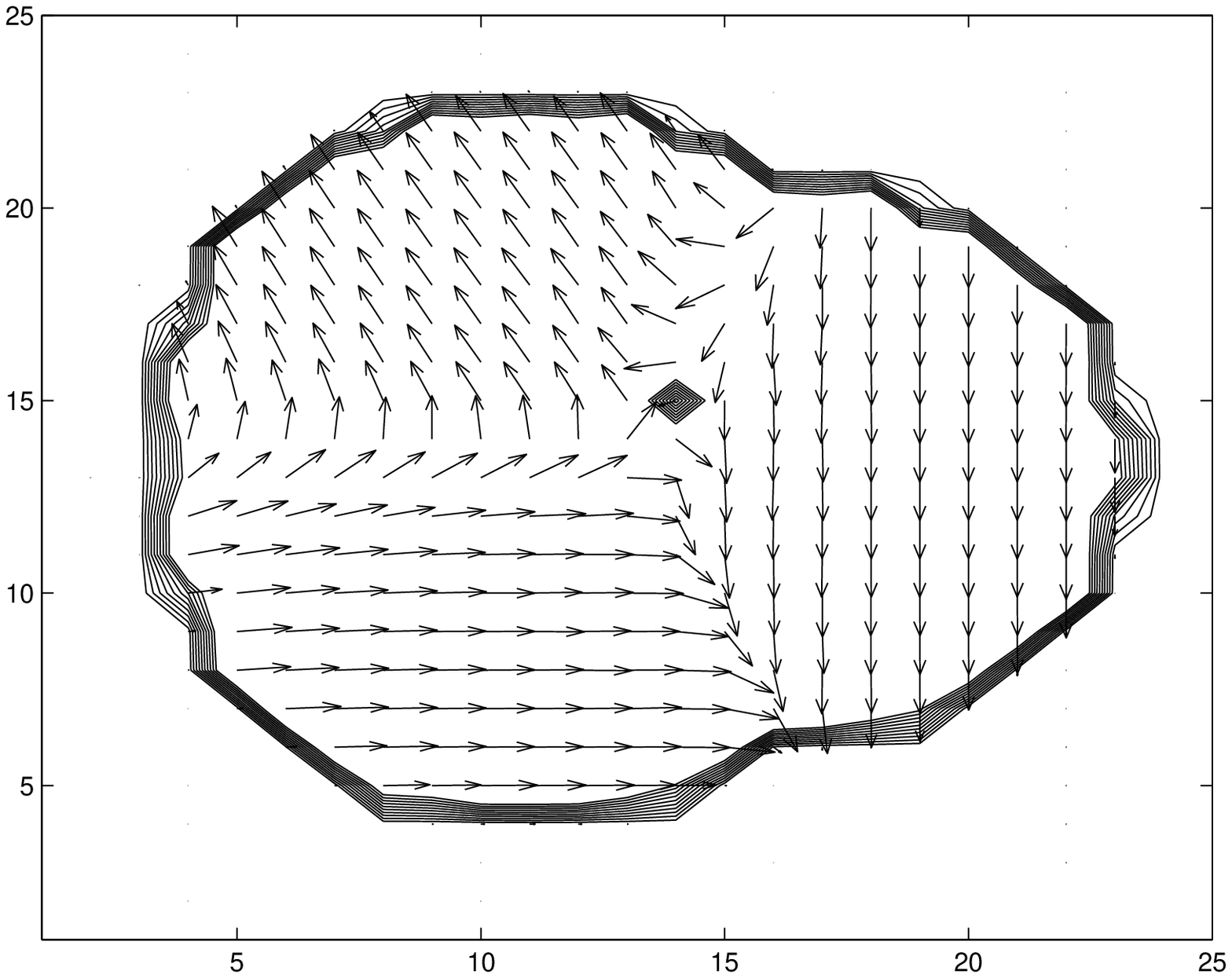,
    width=7.8cm}\hspace*{5mm}\epsfig{figure=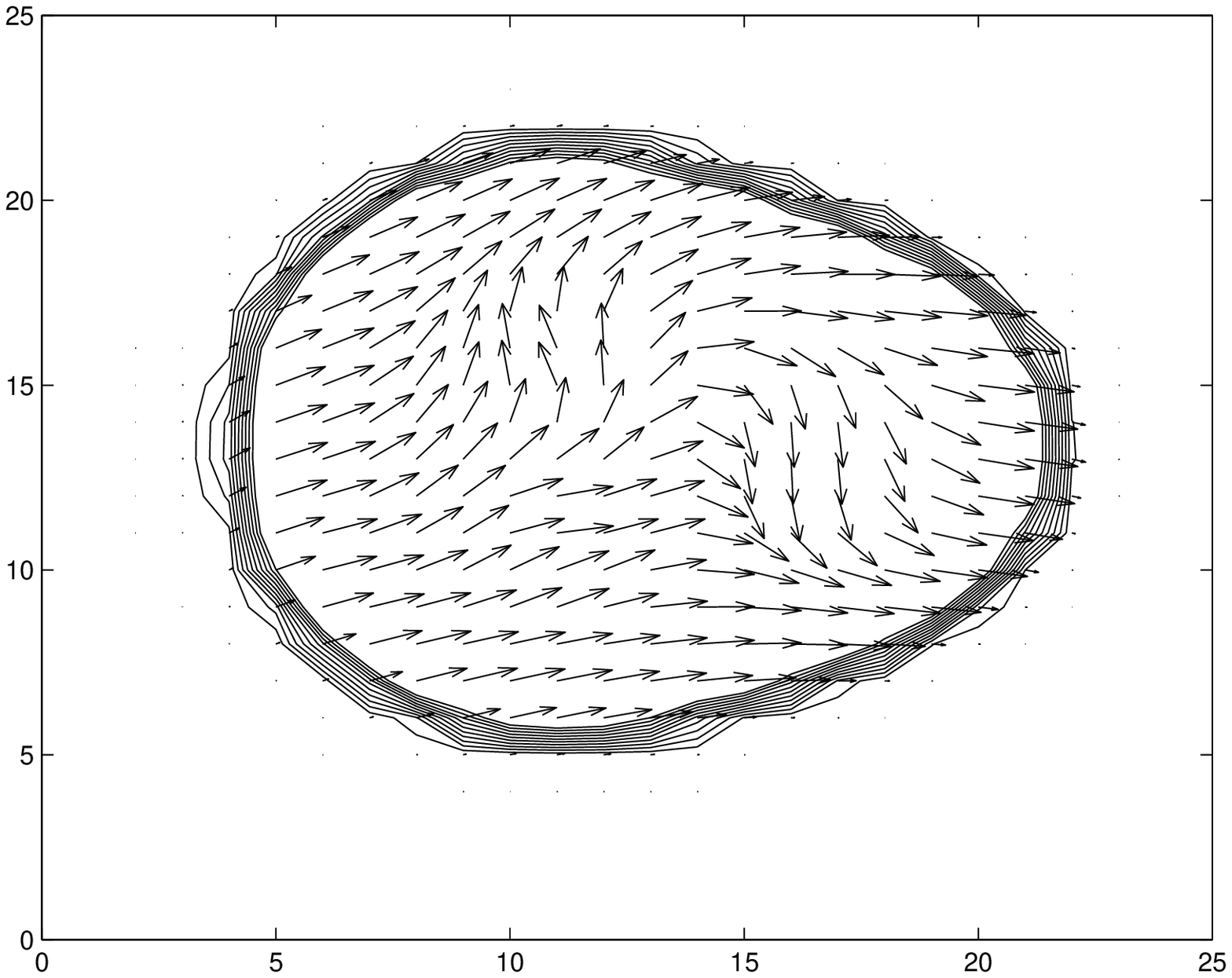,
    width=7.8cm} \lbfig{vortex}
\end{center}
\vspace*{-6.4cm}\noindent \hspace*{1.4cm}{\large\bf a)}
\hspace*{7.85cm}{\large\bf b)} \vspace*{5.4cm}
\begin{center}
\caption{Phase plot and bubble walls after three-bubble collisions,
  with phases 0 (bottom left), $2\pi / 3$ (top left) and $-\pi / 2$
  (right): a) with $\Gamma = 0$ a vortex is formed at the centre, and
  b) with identical initial conditions, but $\Gamma = 0.5$ there is no
  vortex.}
\end{center}
\end{figure*}

Figure \ref{vortex} illustrates our findings -- it shows a
cross-section through a non-simultaneous three-bubble collision, after
all three bubbles have merged.  In each case, the bubbles of initial
radius $R=5$, centred at $(\pm 8,0,-10)$ and $(0,0,10)$, were given
phases $\theta= {-\pi / 2}, {0}$ and ${2 \pi /3}$.  For identical
initial conditions, we see that in the fast-moving case a vortex is
formed, but when the bubbles are slowed down, the phase difference
between the two bubbles has equilibrated by the time the third bubble
collides, and no defect is formed.

In any cosmological phase transition where the bubble wall is
significantly slowed down, we may also expect the plasma to have
non-zero conductivity, which will affect the evolution of the fields
and so needs to be considered in any attempt at a realistic model.  We
have simulated the effects of the finite-conductivity of the plasma,
by adding on to the right-hand side of the gauge field equations
(\ref{local_a}) a conduction current $j_c^\mu$, whose spatial part is
given by
\begin{equation}
{\bf j_c} = \sigma {\bf E}.
\end{equation}
The corresponding charge density $\rho_c$ is fixed by the continuity
relation $\partial_\mu j_c^\mu = 0$.  For large values of the
conductivity, it has been shown \cite{kibble_vilenkin} that the
oscillations in the gauge-invariant phase difference, which took place
in fast-moving bubbles with $\sigma=0$, are exponentially damped.  

Figure \ref{sigma} (a) shows the evolution of the gauge-invariant
phase difference in this case, where the walls are moving at the speed
of light, for three different values of the conductivity $\sigma$.  We
confirm that, as $\sigma$ increases, the phase oscillations are more
heavily suppressed, with practically no oscillations occurring for
$\sigma \gtrsim 0.5$.  In Figure \ref{sigma} (b), we present the
results of our simulations for slow-moving bubbles.  For $\sigma=0$,
we have the case considered above -- heavily suppressed oscillations.
Increasing $\sigma$ merely serves to increase the suppression of phase
oscillations: no new effect is observed.

Whereas it is already known that for bubbles moving at the speed of
light, phase oscillations can be killed by a high conductivity
\cite{kibble_vilenkin}, it is clear from our work that in
slower-moving bubbles, the same effect can be obtained by a much lower
value of $\sigma$.

\begin{figure*}[ht]
\begin{center}
  \epsfig{figure=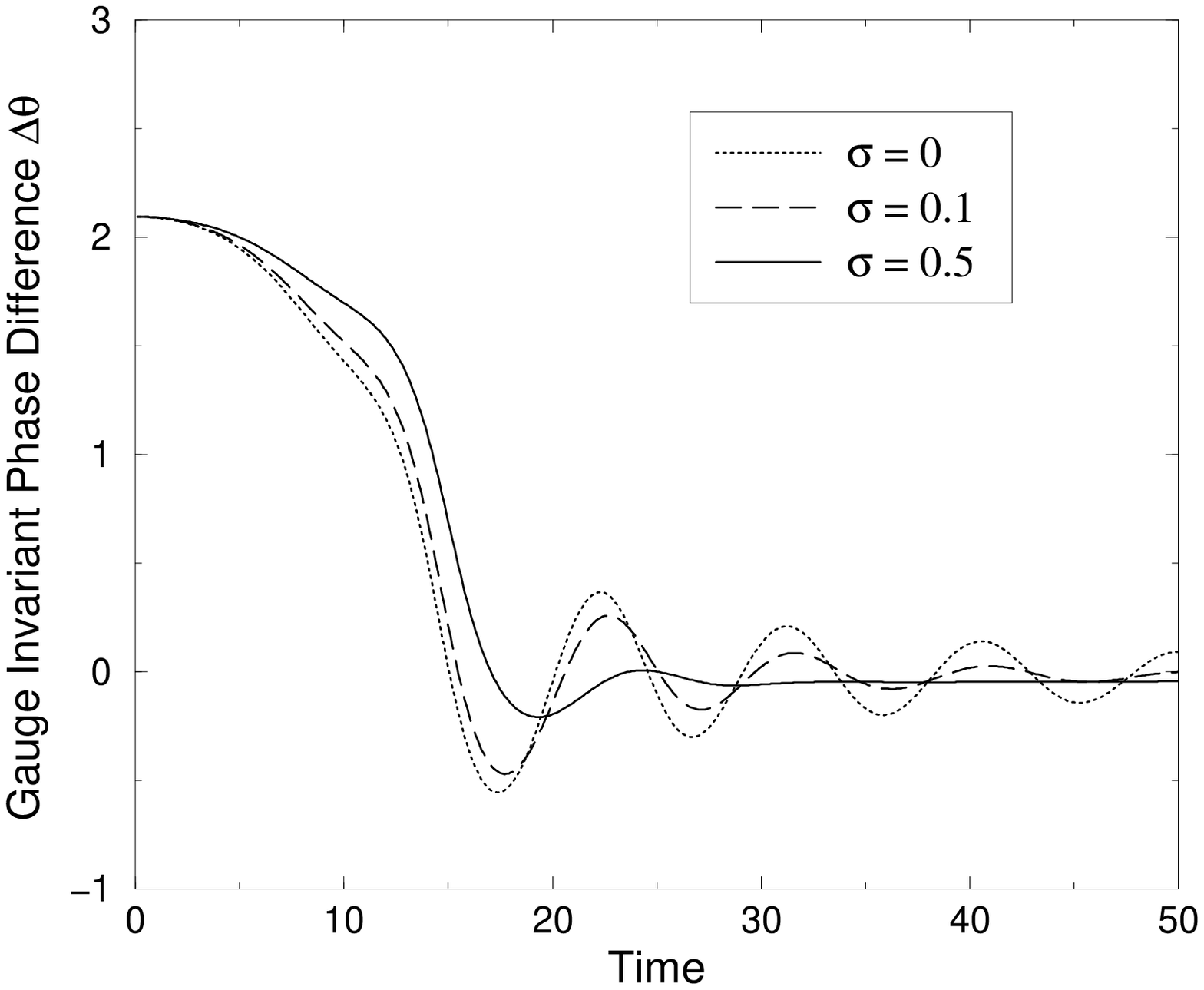,
    width=7.8cm}\hspace*{5mm}\epsfig{figure=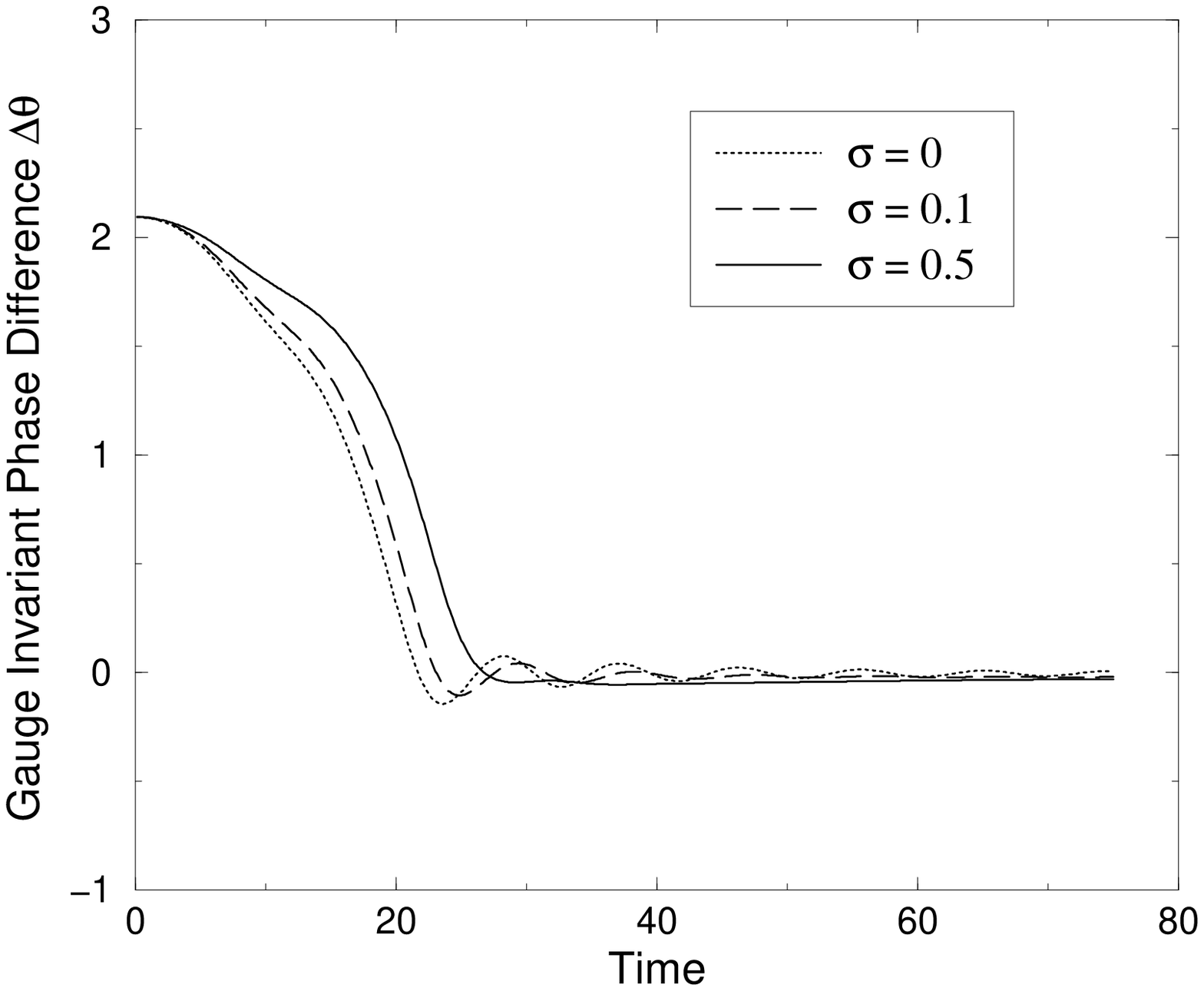,
    width=7.8cm} \lbfig{sigma}
\end{center}
\vspace*{-6.4cm}\noindent \hspace*{1.9cm}{\large\bf a)}
\hspace*{7.85cm}{\large\bf b)} \vspace*{5.4cm}
\begin{center}
\caption{Gauge-invariant phase difference between two bubbles moving
at the speed of light, with conductivity $\sigma = 0, 0.1$ and $0.5$.
In each case, the initial phase difference is $2 \pi / 3$, and $e =
0.5$.  In Figure a) $\Gamma = 0$, and in Figure b) $\Gamma = 0.5$.}

\end{center}
\end{figure*}

\subsection{Extra defect production}

An interesting consequence of slower-moving bubbles concerns the issue
of `extra' defect production at a two-bubble collision.  Hawking, Moss
and Stewart \cite{HMS} first described (by energy considerations) how
two true-vacuum bubbles travelling at nearly the speed of light would
`pass through each other', leading to the temporary restoration of the
spontaneously-broken symmetry in a region between the two bubbles.
This is illustrated in Figure \ref{bounce} -- the two bubbles collide,
and bounce off of (or pass through) each other, producing a region of
$\Phi = 0$ false vacuum inside the merged bubbles, which decays via
oscillations of the bubble walls into the true-vacuum state.  The size
of the symmetry-restored region, and the time taken to decay
completely to the true vacuum depends on the initial phase difference
between the two bubbles, and the asymmetry parameter $\varepsilon$.
Copeland and Saffin \cite{ed_paul} showed how this could lead to the
formation of `extra' -- in the sense that a defect would not be
expected from the initial distribution of phases -- flux-tube vortices
in a gauge theory, around these regions of temporarily restored
symmetry, and hence to an increase in the initial defect density after
a phase transition.

Our simulations show -- see Figure \ref{merge} -- that dissipation
prevents this bouncing, or passing-through, of the bubbles.  The
excess energy, which would cause the symmetry restoration, is
dissipated away by the plasma, and the bubbles simply merge.  Thus
there is no symmetry-restored region around which a non-zero winding
of the phase can occur, and so no `extra' defects would be formed.

\section{Magnetic Fields}

Another consequence of first-order phase transitions which may be
cosmologically significant is the generation of primordial magnetic
fields.  Galaxies are observed to have magnetic fields $B_{\rm gal}
\sim 10^{-6}G$, coherent over large scales.  Given a small initial
seed field a dynamo mechanism, powered by the differential rotation of
the galaxy in combination with the small-scale turbulent motion of the
ionized gas, could generate the observed galactic fields.  Many
mechanisms for producing such a seed field have been proposed, one
being bubble collisions at a first-order phase transition (see e.g.
\cite{ola} and references therein).  It had been believed that it was
not possible to generate a seed field of sufficient magnitude at the
electroweak phase transition for the dynamo mechanism to explain
galactic magnetic fields as large as $10^{-6}G$.  However a recent
paper by Davis, Lilley and T\"{o}rnkvist \cite{DLT} showed how, in a
universe with a low matter density and in particular a positive
cosmological constant, a dynamo mechanism may be able to generate
observed galactic magnetic fields from a much smaller seed field.  As
a consequence, electroweak-scale magnetic fields may be viable
primordial seed fields, and it is of interest to consider the effect
of slow-moving bubble walls and finite plasma conductivity on the
generation of magnetic fields.

If the gauge fields are set to zero initially, it can be seen from the
equations of motion (\ref{local_rho}), (\ref{local_theta}) and
(\ref{local_a}), that non-zero gauge fields can only be generated
where there exist spatial phase gradients, that is after the collision
of two or more bubbles.  After the collision of two bubbles, a loop of
magnetic flux is generated around the circle of intersection
\cite{kibble_vilenkin}.  The amount of flux generated is given by the
integral of the gauge field $A$ around any loop which passes outside
the bubbles.  This is the same for all bubbles, regardless of size or
speed
\begin{equation} \label{flux}
\ointop A_i dx^i = \frac{2 \theta_0}{e},
\end{equation}
and our simulations confirm this.  When a third bubble collides, the
fluxes combine, and if there is a phase winding of $2\pi$ around the
centre, one flux quantum $2\pi / e$ will be trapped.
\begin{figure*}[ht]
\begin{center}
  \epsfig{figure=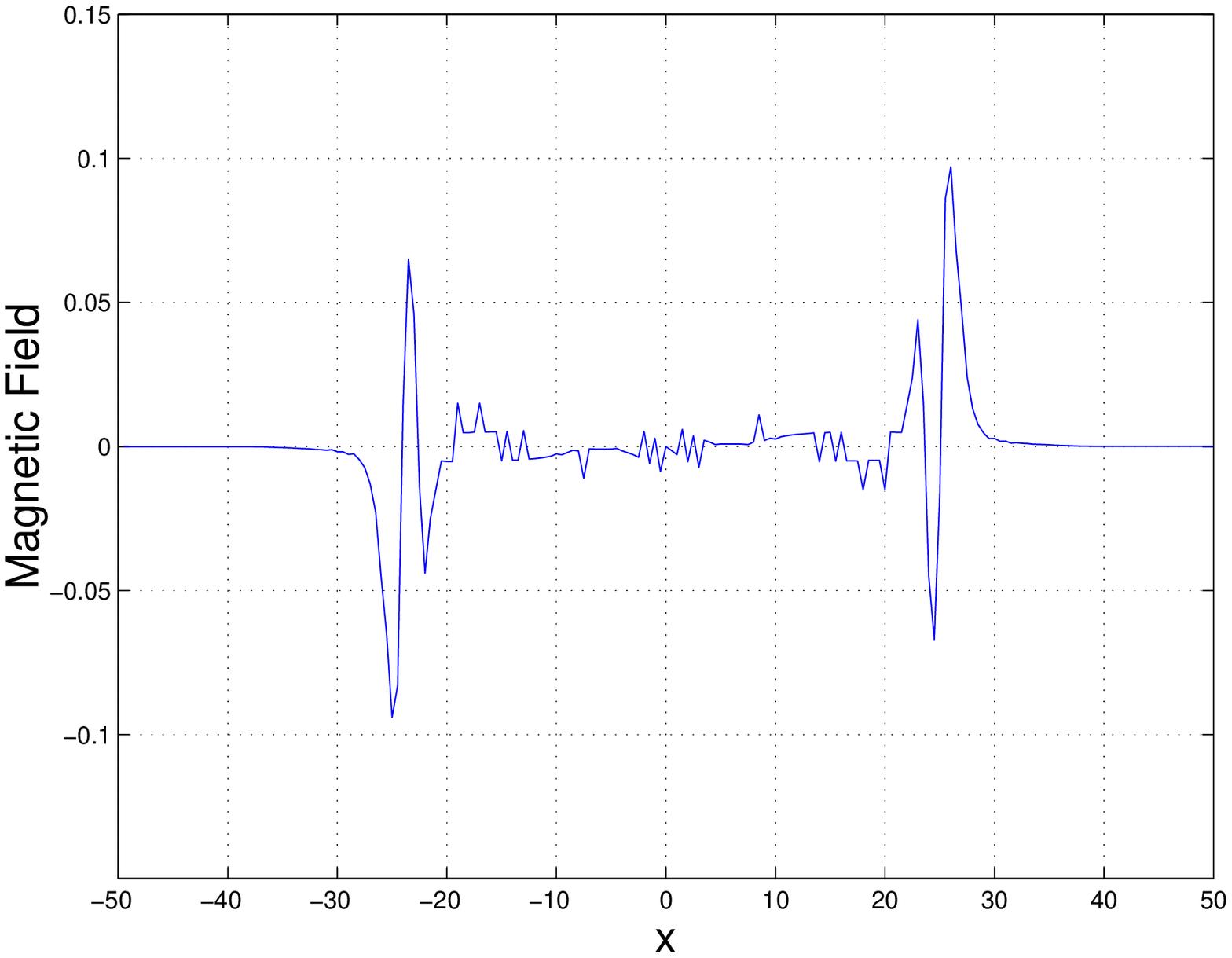,
    width=5.8cm}\hspace*{2mm}\epsfig{figure=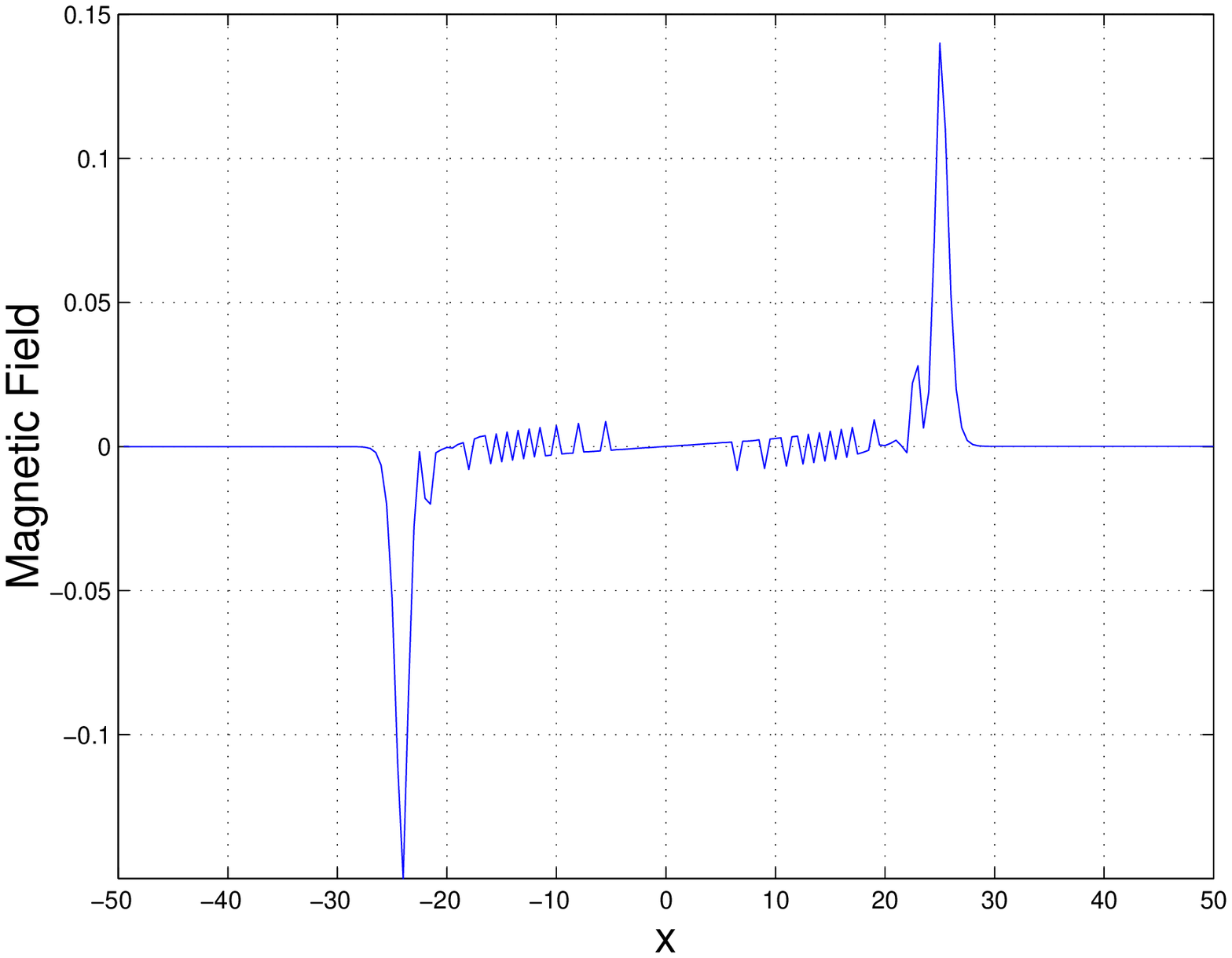,
    width=5.8cm} \hspace*{2mm}\epsfig{figure=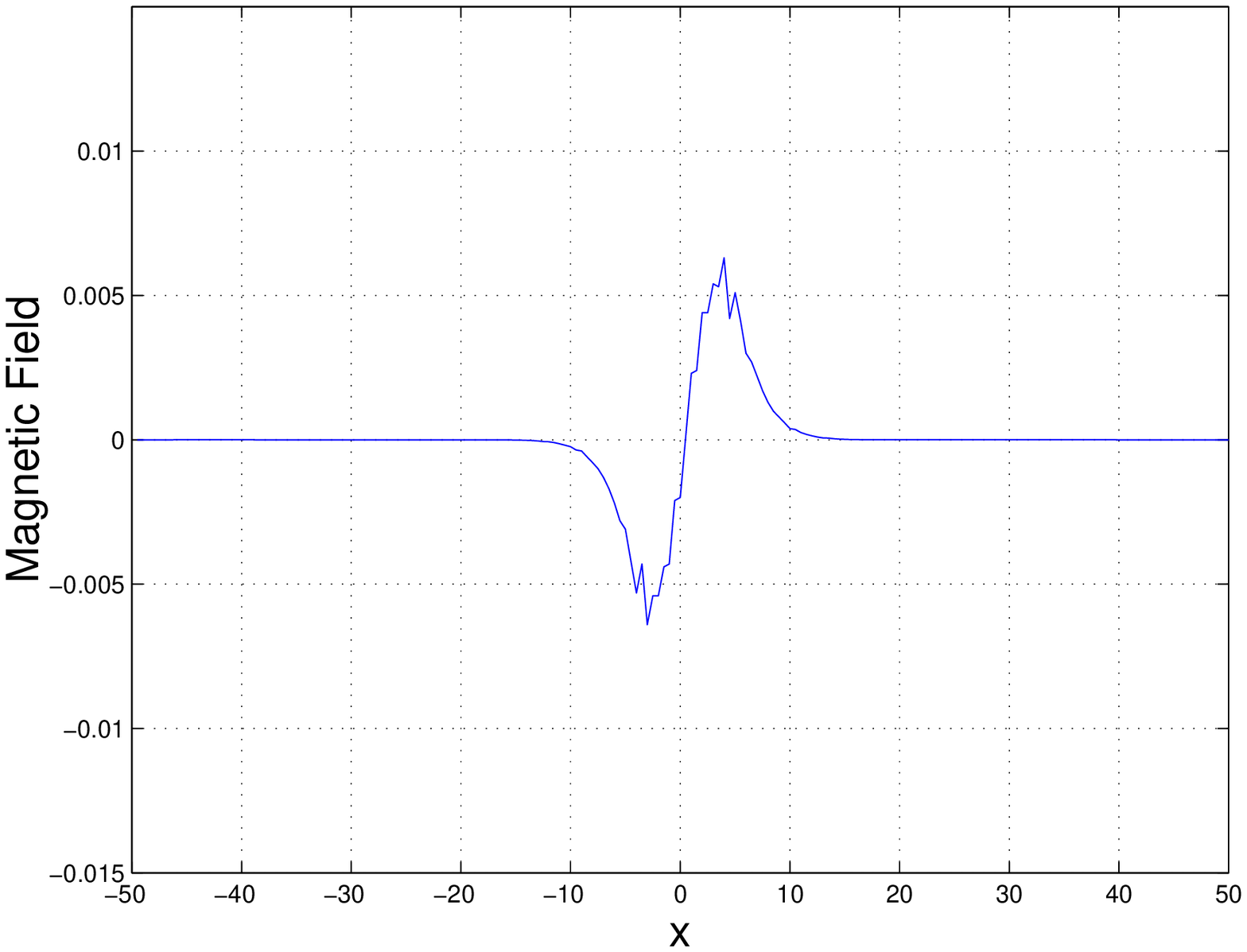,
    width=5.8cm}\lbfig{magfield}
\end{center}
\vspace*{-5.1cm}\noindent \hspace*{0.65cm}{\large\bf a)}
\hspace*{5.45cm}{\large\bf b)}\hspace*{5.7cm}{\large\bf c)}
\vspace*{4cm}
\begin{center}
\caption{Magnetic field strength along radius of collision of two
  bubbles. In figures (a) and (b) the bubbles are expanding at the
  speed of light, and in figure (c) the bubbles are slow-moving, with
  $\Gamma = 2$.  On the left, conductivity $\sigma=0$, in the centre
  $\sigma = 0.5$ and on the right $\sigma = 5$.  The bubbles, of
  initial radius $R_0 = 5$ centred at $(\pm 8,0,0)$, were given phases
  $\theta = 0$ and $\theta = 2 \pi/3$.  The figures show the field
  strength at $t=35$ for the fast-moving and $t=52.5$ for the
  slow-moving bubbles.  In the fast-moving case, with zero
  conductivity, a secondary peak of opposite sign follows inside the
  primary peak, as described in {\protect \cite{CST}}.  For non-zero
  conductivity, this secondary peak does not occur -- all the flux is
  concentrated in the primary peak at the edge of the bubbles.  In the
  slow-moving case, a smaller magnetic field forms which is spread
  over the bubble interior, rather than being concentrated in a peak
  at the edge.}
\end{center}
\end{figure*}

If there is no plasma conductivity, the magnetic flux generated is
free to propagate at the speed of light away from the bubble
collision.  If the bubbles are expanding at the speed of light, then
the fields can disperse no further outwards than the intersection of
the bubbles.  Copeland, Saffin and T\"{o}rnkvist \cite{olaconf},
\cite{CST} demonstrated how in this case {\em two} tubes of flux
are produced -- a `primary flux tube' at the intersection of the
collided bubbles and a smaller, `secondary' peak of opposite direction
following behind it -- see Figure \ref{magfield} (a).  Our simulations
show that when non-zero conductivity is included, this secondary peak
does not occur, and all the magnetic flux is concentrated into the
primary peak, which is consequently larger -- Figure \ref{magfield}
(b).  We might expect that in this case, a larger magnetic field would
form as all the flux generated by the two-bubble collision is aligned.

If, however, the bubbles are moving at speeds less than the speed of
light, the flux is able to disperse into the plasma.  Unless the
bubble nucleation rate is extremely high (when a third bubble might be
expected to collide quickly after the initial collision), no magnetic
field will be able to form. For slow-moving bubbles, large-enough
conductivity prevents the flux from dispersing, freezing it in to the
plasma.  Figure \ref{magfield} (c) shows the magnetic field strength
formed after a collision of two slow-moving bubbles, for $\sigma = 5$
(in reference \cite{kibble_vilenkin} an estimate of $\sigma = T/e^2$
is given.  The temperature at a phase transition is typically $T \sim
\eta$, and so $\sigma = 5$ may be realistic).  It can be seen that for
this value of $\sigma$, a magnetic field {\em does} form, but it is
spread through the inside of the bubble, rather than being
concentrated in one or two narrow peaks at the wall, as seen in the
fast-moving case.  The height of the peak is lower by an order of
magnitude -- this is finite, rather than infinite conductivity and so
some flux still escapes.

We note in passing here that provided that the plasma dynamics which
slow down the bubble walls do not affect the bubble nucleation rate,
the average number of bubbles nucleated per unit volume by the time
the phase transition is completed will increase as the bubble wall
velocity decreases.  That is the average bubble radius on collision,
the correlation length of the Higgs field $\xi$, decreases as the wall
velocity decreases.  Since the amount of flux generated at each
collision is independent of the bubble radius, slow-moving bubbles
will generate {\em more} flux, and hence a larger magnetic field when
coarse-grained over many bubble radii.

\section{Conclusion}

In this paper, we have examined the behaviour of colliding true-vacuum
bubbles at a first-order cosmological phase transition.  In the
Abelian Higgs model, strings may form at the collision of three or
more bubbles, but since a simultaneous three-bubble collision is very
unlikely, the dynamics of the phase inside two-bubble collisions is
crucial -- if phase differences between two bubbles can be
equilibrated quickly, and before the arrival of a third bubble, a
topological defect will not form.

The most relevant phase transitions to cosmology involve gauge fields
coupled to the symmetry-breaking field. In such phase transitions, the
speed of the bubble walls will be considerably less than the speed of
light, and yet the phase dynamics of slow-moving bubbles in a gauge
field have not been considered previously.  We have thus paid
particular attention to the evolution of the phase inside collisions
of bubbles moving at speeds much lower than the speed of light, in a
$U(1)$ gauge theory.

In the simplest model, with no gauge fields and where the bubble walls
accelerate up to the speed of light, the phase difference between two
points is found to equilibrate.  In models with a global symmetry
where the bubble walls move slowly, and models with a local symmetry
where the walls move at the speed of light, decaying phase
oscillations have been observed.  We find that in a $U(1)$ gauge
theory, with slow-moving bubble walls, these oscillations are
suppressed.  On collision of two bubbles, instantaneous phase
equilibration is observed. This would lead to a decrease in the
initial expected defect density compared to the other cases.  We have
illustrated our claims by demonstrating an example of the suppression
of defect formation in a local theory, due to nontrivial phase
dynamics.

When two bubbles collide and merge, there will exist phase gradients
across the intersection, a potential difference.  In local theories it
is necessary to define a gauge-invariant notion of the phase
difference, which involves the gauge fields.  Thus in local theories,
the phase difference may equilibrate through the generation of gauge
fields -- there is in effect an `extra channel' for the decay of the
potential difference created on collision.  This explains why we would
expect fewer defects in local theories than in global ones.  In a
local theory, the phase difference between two bubbles is observed to
equilibrate more quickly in slower-moving bubbles than in bubbles
moving at the speed of light.  This is due to the fact in slow-moving
bubbles the rate of generation of phase gradients is lower, yet the
gauge fields are not restricted to propagate at the speed of the
bubble wall and are thus able to equilibrate the phase difference more
rapidly.

If phase equilibration and hence the suppression of defect formation
is aided by the coupling of gauge fields to the Higgs, it is
interesting to ask whether another scalar field $\chi$ could have the
same effect.  In order to be able to dissipate the potential energy in
the phase gradient such a field would need to couple to the phase of
the Higgs field, but also preserve the U(1) symmetry of the
Lagrangian.  This can only be achieved (with terms at most quadratic
in $\Phi$ and its derivatives) by Higgs couplings proportional to
$\partial_\mu \chi [ \Phi^{\dagger}\partial_{\mu}\Phi -
({\partial_{\mu}\Phi}^{\dagger}) \Phi] = \partial_\mu \chi [ 2 i
\rho^2 \partial_\mu \theta]$, but in this case $\chi$ is effectively a
gauge field (\ref{local_a}).  It is possible that fermion couplings to
the Higgs field would aid phase equilibration, but unfortunately this
cannot be simulated easily.

We predict that fewer defects will form in gauge theories than
global-symmetry theories since the phase difference is non-zero for
less time after collision in gauge theories.  In fact, if the bubble
nucleation rate is low enough, it might be possible to effectively
rule out the formation of defects, solely on phase-dynamical grounds,
though percolation could presumably still be achieved.  This is a very
interesting prospect, which could have significant implications for
cosmology -- it may be possible for example to circumvent the monopole
problem without needing inflation if defect formation is dynamically
suppressed in this way.

It has also been seen how it is unlikely that `extra defects', caused
by bubbles bouncing off of each other on collision, will be formed in
cosmological phase transitions, since the bubbles are retarded
sufficiently by the plasma for no such bouncing to occur.  First-order
phase transitions can also generate a primordial magnetic field, which
may seed the galactic dynamo and hence be responsible for the galactic
magnetic fields observed today.  A simple qualitative analysis
suggests that in fast-moving bubble walls, high conductivity (as would
be expected in the early universe) would lead to the generation of a
larger magnetic field.  Where the bubble walls move slower, we have
demonstrated that a magnetic field {\em can} form if the plasma has
non-zero conductivity.  In this case, the smaller average bubble
radius on collision may cause more flux to be generated, producing a
larger magnetic field.  This may be significant in helping to beat the
lower-bound required by the dynamo model in order to produce observed
fields.

We have shown qualitatively how we expect the defect-formation
probability to be decreased by phase equilibration in two-bubble
collisions.  It would be interesting to perform a statistical
simulation of the type done in \cite{ferrera98}, but for gauge-theory
phase transitions, to see quantitatively how the defect density is
affected by the terminal velocity of the walls or the introduction of
gauge fields.  We note that the argument given in the Introduction for
nucleating bubbles with zero gauge fields does not apply if there
already exists a primordial magnetic field before bubble nucleation.
In this case, it is not at all clear what the preferred nucleation
field configuration would be, and we believe that a study of bubble
nucleation in the presence of a magnetic field would be worth while.
It is also of some interest to consider the effect on phase dynamics
and defect formation of the Hubble expansion, since this acts as a
dissipation term on the phase as well as on the bubble walls.  We
conclude, though, with a summary of our findings.

In gauge theories, more defects are formed by fast-moving bubble walls
than by slower ones.  In global theories, the same is true.

For fast-moving bubble walls, more defects are formed in global
theories than in local ones.  For slow-moving bubble walls, the same
is true.

\section{Acknowledgements}
We would like to thank T.Kibble, P.Saffin, D.Steer and especially O.
T\"{o}rnkvist for helpful comments and conversations.  Computer
facilities were provided by the UK National Cosmology Supercomputing
Centre in cooperation with Silicon Graphics/Cray Research, supported
by HEFCE and PPARC.  This work was supported in part by PPARC and an
ESF network grant.  Support for M.L. was provided by a PPARC
studentship and Fitzwilliam College, Cambridge.

\end{document}